\documentclass[sigconf]{acmart}
\renewcommand\footnotetextcopyrightpermission[1]{} 
\usepackage{bm}
\usepackage{amsmath}
\usepackage{amssymb}
\usepackage{amsfonts}

\usepackage{wrapfig}
\usepackage{url}
\usepackage{comment}
\usepackage{graphicx}  
\usepackage{caption}
\usepackage{subcaption}
\usepackage{bbm}
\usepackage{multirow}

\AtBeginDocument{%
  \providecommand\BibTeX{{%
    \normalfont B\kern-0.5em{\scshape i\kern-0.25em b}\kern-0.8em\TeX}}}

\begin{document}
\title{Deep Learning for Insider Threat Detection: Review, Challenges and Opportunities}
\author{Shuhan Yuan}
\affiliation{Utah State University \\ Logan, UT, USA}
\email{shuhan.yuan@usu.edu}

\author{Xintao Wu}
\affiliation{University of Arkansas\\ Fayetteville, AR, USA}
\email{xintaowu@uark.edu}

\begin{abstract}
Insider threats, as one type of the most challenging threats in cyberspace, usually cause significant loss to organizations. While the problem of insider threat detection has been studied for a long time in both security and data mining communities, the traditional machine learning based detection approaches, which heavily rely on feature engineering, are hard to accurately capture the behavior difference between insiders and normal users due to various challenges related to the characteristics of underlying data, such as high-dimensionality, complexity, heterogeneity, sparsity, lack of labeled insider threats, and the subtle and adaptive nature of insider threats.  Advanced deep learning techniques provide a new paradigm to learn end-to-end models from complex data. In this brief survey, we first introduce  one commonly-used dataset for insider threat detection and review the recent literature about deep learning for such research. The existing studies show that compared with traditional machine learning algorithms, deep learning models can improve the performance of insider threat detection. However, applying deep learning to further advance the insider threat detection task still faces several limitations, such as lack of labeled data, adaptive attacks. We then discuss such challenges and suggest future research directions that have the potential to address challenges and further boost the performance of deep learning for insider threat detection. 
\end{abstract}


\maketitle
\pagestyle{empty}

\section{Introduction}

Insider threats are malicious threats from people within the organization, which usually involve intentional fraud, the theft of confidential or commercially valuable information, or the sabotage of computer systems.
The subtle and dynamic nature of insider threats makes detection extremely difficult. The 2018 U.S. State of Cybercrime Survey indicates that 25\% of the cyberattacks are committed by insiders, and 30\% of respondents indicate incidents caused by insider attacks are more costly or damaging than outsider attacks \cite{cso2018StateCybercrime2018}. 

According to the latest technical report~\cite{costaInsiderThreatIndicator2016} from the CERT Coordination Center (CERT/CC), a malicious insider is defined as ``a current or former employee, contractor, or business partner who has or had authorized access to an organization’s network, system, or data, and has intentionally exceeded or intentionally used that access in a manner that negatively affected the confidentiality, integrity, or availability of the organization’s information or information systems.'' 

Compared to the external attacks whose footprints are difficult to hide, the attacks from insiders are hard to detect because the malicious insiders already have the authorized power to access the internal information systems. In general, there are three types of insiders, i.e., \textit{traitors} who misuse their privileges to commit malicious activities, \textit{masqueraders} who conduct illegal actions on behalf of legitimate employees of an institute, and \textit{unintentional perpetrators} who unintentionally make mistakes \cite{liuDetectingPreventingCyber2018}. Based on the malicious activities conducted by the insiders, the insider threats can also be categorized into three types, \textit{IT sabotage} which directly uses IT to make harm to an institute, \textit{theft of intellectual property} which steals information from the institute, \textit{fraud} which indicates unauthorized modification, addition, or deletion of data \cite{homoliakInsightInsidersIT2019a}. 

Insider threat detection has attracted significant attentions over the last decade, where various insider threat detection approaches have been proposed \cite{eldardiryMultiDomainInformationFusion2013,rashidNewTakeDetecting2016,leEvaluatingInsiderThreat2018,salemSurveyInsiderAttack2008,sanzgiriClassificationInsiderThreat2016,tuorDeepLearningUnsupervised2017}. Most of existing approaches achieve the insider threat detection by analyzing the users' behaviors via the audit data, such as host-based data that record activities of users on their own computers, network-based data that are recorded by network equipments, and context data that record users' profile information. A recent survey \cite{sanzgiriClassificationInsiderThreat2016} further categorizes the insider threat detection techniques into 9 classes based on strategies and features used in detection: (1) anomaly based approaches, (2) role based access control, (3) scenario based techniques, (4) decoy documents and honeypot techniques, (5) risk analysis using psychological factors, (6) risk analysis using workflow, (7) improving defense of the network, (8) improving defense by access control, and (9) process control to dissuade insiders. 


Although existing approaches demonstrate good performance on insider threat detection, the traditional shallow machine learning models are unable to make full use of the user behavior data due to their high-dimensionality, complexity, heterogeneity, and sparsity. On the other hand, deep learning as a representation learning algorithm, which is able to learn multiple levels of hidden representations from the complicated data based on its deep structure \cite{bengioDeepLearningRepresentations2013a,lecunDeepLearning2015}, can be used as a powerful tool to analyze the user behavior in an organization to identify the potential malicious activities from insiders. 

Recently, by leveraging the deep feedforward neural network, convolutional neural network (CNN), recurrent neural network (RNN), and graph neural network (GNN), several approaches have been proposed for insider threat detection \cite{liuAnomalyBasedInsiderThreat2018,luInsiderThreatDetection2019,jiangAnomalyDetectionGraph2019,huInsiderThreatDetection2019}. For example, some RNN-based models are proposed to analyze the user sequential data to identify the malicious activities \cite{luInsiderThreatDetection2019,yuanInsiderThreatDetection2019,zhangRolebasedLogAnalysis2018}, while a GNN-based model is investigated to detect the insiders based on the user structural data in an organization \cite{jiangAnomalyDetectionGraph2019}. However, using deep learning for insider threat detection still faces various challenges related to the characteristics of insider threat detection data, such as extremely small number of malicious activities and adaptive attacks. Hence, developing advanced deep learning models that can improve the performance of insider threat detection is still under-explored.

Currently, there is no existing review on the topic of deep learning for insider threat detection. We do not aim to provide a comprehensive survey on the domain of insider threat (see \cite{homoliakInsightInsidersIT2019a,liuDetectingPreventingCyber2018} for details) or a general review on deep learning (see \cite{lecunDeepLearning2015}). In this work, we focus on reviewing the current progresses and pointing out potential future directions of deep learning for insider threat detection. We first briefly review the deep learning and its application on anomaly detection in Section \ref{sec:deep learning}. In Section \ref{sec:review} we introduce one commonly-used dataset used in insider threat detection,  explain why deep learning is needed for insider threat detection, and review a few recent research works of deep learning based insider threat detection. In Section \ref{sec:challenge}, we identify and categorize challenges into extremely unbalanced data, subtle attacks, temporal information in attacks, heterogeneous data fusion, adaptive threats, fine-grained detection, early detection, explainability, lack of testbed, and lack of practical metrics. In Section \ref{sec:future}, we point out research opportunities of insider threat detection based on few-shot learning, self-supervised learning, deep marked temporal point process, multi-modal learning, deep survival analysis, deep Bayesian nonparametric learning, deep reinforcement learning, explainable deep learning in addition to testbed development.  Finally we conclude this survey in Section \ref{sec:conclusion}. 
\section{Deep Learning and its Application on Anomaly Detection}
\label{sec:deep learning}

\subsection{Deep Learning}
With great achievement in various domains, such as computer vision \cite{heDeepResidualLearning2016}, natural language process \cite{liuSurveyContextualEmbeddings2020}, and speech recognition \cite{dengNewTypesDeep2013}, deep learning has dominated the machine learning community in the past few years \cite{lecunDeepLearning2015}. 
Compared with traditional machine learning models that heavily rely on hand engineering to identify  useful features to represent raw data, deep learning models are able to learn semantic representations from the raw data with minimal human efforts. The deep learning models as representation learning models adopt a multi-layer structure to learn data representation, where the lower layers capture the low-level features of data, while the higher layers extract the high-level abstract. 

Deep learning models can be broadly categorized into four groups based on their architectures: (1) deep feedforward neural network (DFNN), which includes a number of deep learning models consisting of multiple layers, such as deep belief network \cite{hintonFastLearningAlgorithm2006}, deep Boltzmann machine \cite{salakhutdinovDeepBoltzmannMachines2009} and deep autoencoder \cite{vincentExtractingComposingRobust2008}; (2) convolutional neural network (CNN), which leverages the convolutional and pooling layers to achieve the shift-invariant property; (3) recursive neural network (RvNN), which takes a recursive data structure of variable sizes and makes predictions in a hierarchical structure; (4) recurrent neural network (RNN), which maintains an internal hidden state to capture the sequential information. Since the deep learning field is growing so fast, many new types of deep structures are proposed each year. The readers can refer to recent surveys, e.g., \cite{arulkumaranDeepReinforcementLearning2017,chalapathyDeepLearningAnomaly2019,pouyanfarSurveyDeepLearning2018,zhangDeepLearningSentiment2018,zhangDeepLearningGraphs2018}, to learn more information about deep learning and its applications in various domains.


\subsection{Deep Learning for Anomaly Detection}
Anomaly detection is to identify instances that are dissimilar to all others, which is an important problem with multiple applications, such as fraud detection, intrusion detection, and video surveillance \cite{chandolaAnomalyDetectionSurvey2009}.
Anomalies are  referred to as abnormalities, deviants, or outliers in the data mining and statistics literature and roughly speaking insider threats can be treated as one type of anomalies. A variety of machine learning and deep learning based anomaly detection approaches have been developed, however, they are not necessarily applicable for insider threat detection due to the characteristics of insider threat as will be shown in Section \ref{sec:review}. A recent survey \cite{chalapathyDeepLearningAnomaly2019} categorizes the deep learning-based anomaly detection into three groups based on the availability of labels, i.e., supervised, semi-supervised, and unsupervised deep anomaly detection. In some cases, when both normal and anomalous data are available, supervised deep anomaly detection approaches are proposed for binary or multi-class classification \cite{chalapathyBidirectionalLSTMCRFClinical2016,chalapathyInvestigationRecurrentNeural2016}. A more common scenario is that it is easy to collect many normal samples while only a small number of anomalous samples is available, so that semi-supervised deep anomaly detection can be adopted by leveraging the normal samples to separate outliers \cite{songHybridSemiSupervisedAnomaly2017a,zhengOneClassAdversarialNets2019,akcayGANomalySemiSupervisedAnomaly2018}. When no labeled data are available, unsupervised deep anomaly detection is applied to detect anomalies based on intrinsic properties of data samples\cite{tuorDeepLearningUnsupervised2017,hendrycksDeepAnomalyDetection2018,suRobustAnomalyDetection2019}. 

Although many deep learning-based approaches for anomaly detection have been proposed, the performance improvement of deep learning for anomaly detection may not be as significant as deep learning for other domains, such as computer vision and natural language process. The main potential reason is that deep learning models consisting of millions of parameters require a large amount of labeled data for training properly. However, for anomaly detection, it is very difficult, if not impossible, to collect a large number of labeled anomalies in the training data. 
\section{Literature Review}
\label{sec:review}
Insider threats as one of the most challenging attacks to address in cyberspace have attracted much attention for a long time. Various insider threat detection approaches, such as hidden Markov model (HMM) and support vector machine (SVM), have been proposed in literature \cite{homoliakInsightInsidersIT2019a,sanzgiriClassificationInsiderThreat2016}. Moreover, leveraging the deep learning models for insider threat detection is not well-explored in literature as only a few papers are available. It is natural to question the need of deep learning models for insider threat detection. In this section, we first introduce a widely used insider threat dataset and describe its characteristics. We then present why deep learning based insider threat detection models are needed. At the end of this section, we review the existing deep learning based insider threat detection papers in literature and categorize them based on the adopted deep learning architectures.

\begin{table*}[t]
\centering
\caption{Activity Types in Log Files}
\label{tb:types}
\begin{tabular}{|c|l|}
    \hline
    Files                       & \multicolumn{1}{c|}{Operation Types}                                                   \\ \hline
    \multirow{4}{*}{logon.csv}  & Weekday Logon (employee logs on a computer on a weekday at work hours)                 \\ \cline{2-2}
                                & Afterhour Weekday Logon (employee logs on a computer on a weekday after work hours)    \\ \cline{2-2}
                                & Weekend Logon (employees logs on at weekends)                                          \\ \cline{2-2}
                                & Logoff (employee logs off a computer)                                                  \\ \hline
    \multirow{4}{*}{email.csv}  & Send Internal Email (employee sends an internal email)                                 \\ \cline{2-2}
                                & Send External Email (employee sends an external email)                                 \\ \cline{2-2}
                                & View Internal Email (employee views an internal email)                                 \\ \cline{2-2}
                                & View external Email (employee views an external email)                                 \\ \hline
    \multirow{3}{*}{http.csv}   & WWW Visit (employee visits a website)                                                  \\ \cline{2-2}
                                & WWW Download (employee downloads files from a website)                                 \\ \cline{2-2}
                                & WWW Upload (employee uploads files to a website)                                       \\ \hline
    \multirow{4}{*}{device.csv} & Weekday Device Connect (employee connects a device on a weekday at work hours)         \\ \cline{2-2}
                                & Afterhour Weekday Device Connect (employee connects a device on a weekday after hours) \\ \cline{2-2}
                                & Weekend Device Connect (employee connects a device at weekends)                        \\ \cline{2-2}
                                & Disconnect Device (employee disconnects a device)                                      \\ \hline
    \multirow{4}{*}{file.csv}   & Open doc/jpg/txt/zip File (employee opens a doc/jpg/txt/zip file)                      \\ \cline{2-2}
                                & Copy doc/jpg/txt/zip File (employee copies a doc/jpg/txt/zip file)                     \\ \cline{2-2}
                                & Write doc/jpg/txt/zip File (employee writes a doc/jpg/txt/zip file)                    \\ \cline{2-2}
                                & Delete doc/jpg/txt/zip File (employee deletes a doc/jpg/txt/zip file)                  \\ \hline
    \end{tabular}
\end{table*}

\begin{table}[]
    \centering
    \caption{The Statistics of CERT Datasets r4.2 and r6.2}
    \label{tb:cert}
        \begin{tabular}{|c|c|c|c|c|}
            \hline
                & \# employees & \# insiders & \# activities & \begin{tabular}[c]{@{}c@{}}\# malicious \\ activities\end{tabular} \\ \hline
            r4.2 & 1000        & 70          & 32,770,227    & 7323                                                               \\ \hline
            r6.2 & 2500        & 5           & 135,117,169   & 470                                                                \\ \hline
        \end{tabular}
    \end{table}

\subsection{CERT Insider Threat Dataset}
\label{sec:data}

The authors, in the recent survey \cite{homoliakInsightInsidersIT2019a}, present case studies of incidents of insider threat and existing datasets gathered from laboratory experiments and the real world. They further group commonly used eleven datasets into five categories: masquerader-based, traitor-based, miscellaneous malicious, substituted masqueraders, and identification/authentication-based. We refer readers to this survey \cite{homoliakInsightInsidersIT2019a} for detailed information on this topic. However, there is no comprehensive real world dataset publicly available  for insider threat detection.  
Most of the recent studies adopt the synthetic CMU CERT datasets \cite{glasserBridgingGapPragmatic2013} to evaluate their proposed approaches. In this survey, we introduce the CERT dataset to illustrate challenges for insider threat detection. 

The CERT division of Software Engineering Institute at Carnegie Mellon University maintains  a database of more than 1000 real case studies of insider threat and has generated a collection of synthetic insider threat datasets using scenarios containing traitor instances and masquerade activities. 

CERT dataset consists of five log files that record the computer-based activities for all employees in a simulated organization, including \textbf{logon.csv} that records the logon and logoff operations of all employees, \textbf{email.csv} that records all the email operations (send or receive), \textbf{http.csv} that records all the web browsing (visit, download, or upload) operations, \textbf{file.csv} that records activities (open, write, copy or delete) involving a removable media device, and \textbf{decive.csv} that records the usage of a thumb drive (connect or disconnect). Table \ref{tb:types} lists the activity types in each log file. For each activity, it also contains related descriptions. For example, the activity type ``Send Internal Email'' includes time, sender, receivers, subject, and content information. Besides the employees' activity data on computers, the CERT dataset also provides the psychometric score for each employee, known as ``Big Five personality traits'', in \textbf{psychometric.csv}. The data generation process reflects  practical constraints and employs a number of different model types, e.g., graph models of the population’s social structure, topic models for generating content, psychometric models for dictating behavior and preferences, models workplace behaviors.

\begin{table*}
\centering
\caption{The numbers of activities, malicious activities, sessions, and malicious sessions for each insider}
\label{tb:statistic_of_insider}
\begin{tabular}{|l|c|c|c|c|c|}
\hline
                      & ACM2278 & CDE1846 & CMP2946 & MBG3183 & PLJ1771 \\ \hline
Activity \#           & 31370   & 37754   & 61989   & 42438   & 20964   \\ \hline
Malicious activity \# & 22      & 134     & 242     & 4       & 18      \\ \hline
Session \#            & 316     & 374     & 627     & 679     & 770     \\ \hline
Malicious session \#  & 2       & 9       & 53      & 1       & 3       \\ \hline
\end{tabular}
\end{table*}

There are several versions of datasets according to when the datasets were created. The most widely-used versions are r4.2 and r6.2. Table \ref{tb:cert} shows the statistics of these two datasets. In short, r4.2 is a ``dense'' dataset that contains many insiders and malicious activities, while r6.2 is a ``sparse'' dataset that contains 5 insiders and 3995 normal users. For each user, it records activities from January 2010 to June 2011. On average, the number of activities for each employee is around 40000. Specifically, the CERT in r6.2 dataset simulates the following five scenarios of attacks from insiders. 

\begin{itemize}
    \item User ACM2278 who did not previously use removable drives or work after hours begins logging in after hours, use a removable drive, and upload data to wikileaks.org.
    \item User CMP2946 begins surfing job websites and soliciting employment from a competitor. Before leaving the company, they use a thumb drive to steal data.
    \item System administrator PLJ1771 becomes disgruntled. Downloads a keylogger and uses a thumb drive to transfer it to his supervisor's machine. The next day, he uses the collected keylogs to log in as his supervisor and send out an alarming mass email, causing panic in the organization. He leaves the organization immediately.
    \item User CDE1846 logs into another user's machine and searches for interesting files, emailing to their home email.
    \item User MBG3183, as a member of a group decimated by layoffs, uploads documents to Dropbox, planning to use them for personal gain.
\end{itemize}

Table \ref{tb:statistic_of_insider} shows the statistics of each of the above five insiders, including the numbers of activity, malicious activity, session, and malicious sessions.  In sum, insider threat detection is a task like looking for a needle in a haystack, thus it is generally infeasible to manually define features or use shallow machine learning models to detect insider threats. 

\subsection{Why Deep Learning for Insider Threat Detection?}

Among the many attractive properties of deep learning models, the potential advantages of deep learning for insider threat detection can be summarized as follows.

\begin{itemize}
    \item {\bf Representation Learning}. The most significant advantage of deep learning models is based on the capability to automatically discover the features needed for detection. The user behavior in cyberspace is complicated and non-linear. Manually designed features are hard and inefficient to capture user behavior information. Meanwhile, the learning models with shallow-structures, such as HMM and SVM, are relatively simple structures with only one layer for transforming the raw feature into a high-level abstract that can be used for detection. These shallow models are effective for solving many well-constrained problems, but shallow models with limited capability are hard to model complicated user behavior data. Comparatively, deep learning models are able to leverage deep non-linear modules to learn the representation by using a general-purpose learning procedure. Hence, it is a natural fit to use deep learning models to capture complex user behavior and precisely detect  user's intentions, especially those malicious ones.
    
    \item {\bf Sequence Modeling.} Deep learning models, such as recurrent neural network (RNN) and the newly-proposed Transformer, have shown promising performance in modeling sequential data, like video, text, and speech \cite{gravesGeneratingSequencesRecurrent2013,vaswaniAttentionAllYou2017a}. Since it is natural to represent the user activities recorded in audit data as sequential data, leveraging RNN or Transformer to capture the salient information of complicated user behavior has the great potential to boost the performance of insider threat detection.
    
    \item {\bf Heterogeneous Data Composition.} Deep learning models also have achieved great performance on tasks that fuse heterogeneous data, such as image captioning \cite{chengHierarchicalMultimodalAttentionbased2017,huangMiSTMultiviewMultimodal2019}. For insider threat detection, besides modeling the user activity data as sequences, other information, such as user profile information and user structure information in an organization, is also critical. Combining all the useful data for insider threat detection is expected to achieve better performance than only using a single type of data. Compared with traditional machine learning methods, deep learning models are more powerful to combine the heterogeneous data for detection.
\end{itemize}

\subsection{Deep Learning For Insider Threat Detection}
In this subsection, we review the main literature and categorize deep learning based insider threat detection papers based on the adopted deep learning architectures. Table \ref{tb:summary} summarizes all the papers discussed in this section. 

\begin{table*}[]
    \caption{Categorization of Deep Learning based Insider Threat Detection Papers Discussed in this Section}
    \label{tb:summary}
    \begin{tabular}{|c|c|c|c|c|c|}
    \hline
    \multirow{2}{*}{Model}                                                                      & \multirow{2}{*}{Paper}                                   & \multirow{2}{*}{Training} & \multicolumn{3}{c|}{Granularity} \\ \cline{4-6} 
                                                                                                &                                                          &                           & Insider  & Session  & Activity  \\ \hline
    \multirow{2}{*}{\begin{tabular}[c]{@{}c@{}}Deep Feed-forward\\ Neural Network\end{tabular}} & \cite{liuAnomalyBasedInsiderThreat2018} & Unsupervised              &          & $\surd$   &           \\ \cline{2-6} 
                                                                                                & \cite{linInsiderThreatDetection2017}    & One-class                 &          & $\surd$   &           \\ \hline
    \multirow{5}{*}{Recurrent Neural Network}                                                   & \cite{luInsiderThreatDetection2019}     & Unsupervised              &          & $\surd$   &           \\ \cline{2-6} 
                                                                                                & \cite{tuorDeepLearningUnsupervised2017} & Unsupervised              &          & $\surd$   &           \\ \cline{2-6} 
                                                                                                & \cite{zhangRolebasedLogAnalysis2018}    & Unsupervised              &          & $\surd$   &           \\ \cline{2-6} 
                                                                                                & \cite{yuanInsiderThreatDetection2019}   & Unsupervised              &          & $\surd$   &           \\ \cline{2-6} 
                                                                                                & \cite{yuanInsiderThreatDetection2018}   & Supervised                &          & $\surd$   &           \\ \hline
    Convolutional Neural Network                                                                & \cite{huInsiderThreatDetection2019}     & Supervised                &          & $\surd$   &           \\ \hline
    \multirow{2}{*}{Graph Neural Network}                                                       & \cite{jiangAnomalyDetectionGraph2019}   & Supervised                & $\surd$  &           &           \\ \cline{2-6} 
                                                                                                & \cite{liuLog2vecHeterogeneousGraph2019} & Unsupervised              &          &           & $\surd$   \\ \hline
    \end{tabular}
    \end{table*}

As shown in Table \ref{tb:summary}, due to the extremely unbalanced nature of the dataset, most of the proposed approaches adopt the unsupervised learning paradigm for insider threat detection. For the detection granularity, most of the papers focus on detecting malicious subsequence (e.g., activities in 24 hours) or malicious session. Here, a session indicates a sequence of activities among ``Logon'' and ``Logoff''. If there are malicious activities in a session (subsequence), the session (subsequence) will be labeled as a malicious session (subsequence). Due to the limited information that can be leveraged, detecting malicious activities is difficult. Currently, there is only one work focusing on activity level detection. 


\subsubsection{Deep Feedforward Neural Network}

Deep feedforward neural network is a classical type of deep learning model. Various deep learning models are feed-forward neural networks, such as deep autoencoder, deep belief network, and deep Boltzmann machine \cite{pouyanfarSurveyDeepLearning2018}. These deep neural networks are able to learn different levels of representations from the input data based on the multi-layer structures. 

Several studies have proposed to use the deep feadforward neural network for insider threat detection. \cite{liuAnomalyBasedInsiderThreat2018} uses deep autoencoder to detect the insider threat. Deep autoencoder consists of an encoder and a decoder, where the encoder encodes the input data to hidden representations while the decoder aims to reconstruct the input data based on the hidden representations. The objective of the deep autoencoder is to make the reconstructed input close to the original input. Because the majority of activities in an organization are benign, the input with insider threats should have relatively high reconstruction errors. As a result, the reconstruction error of the deep autoencoder can be used as an anomalous score to identify the insider threat. Another idea of leveraging the autoencoder structure is that after learning the hidden representations based on the reconstruction error, a one-class classifier, such as one-class SVM, is applied on the learned hidden representations to identify the insider threats \cite{linInsiderThreatDetection2017}. 

\subsubsection{Recurrent Neural Network}

Recurrent neural network (RNN) is mainly used for modeling the sequential data, which maintains a hidden state with a self-loop connection to encode the information from a sequence \cite{gravesGeneratingSequencesRecurrent2013}. The standard RNN is difficult to train over long sequences due to the vanishing or exploding gradient problem \cite{bengioLearningLongtermDependencies1994}. Currently, two variants of the standard RNN, Long Short-Term Memory (LSTM) \cite{sepphochreiterLongShorttermMemory1997} and Gated Recurrent Unit (GRU) \cite{chungGatedFeedbackRecurrent2015}, are widely-used to model the long sequences and capture the long-time dependence by incorporating gate mechanisms.

The user activities on a computer can be naturally modeled as sequential data. As a result, many RNN based approaches have been proposed to model the user activities \cite{luInsiderThreatDetection2019,tuorDeepLearningUnsupervised2017,yuanInsiderThreatDetection2018,yuanInsiderThreatDetection2019} for insider threat detection. The basic idea is to train an RNN model to predict a user’s next activity or period of activities. As long as the prediction results and the user's real activities do not have significant differences, we consider the user follows the normal behavior. Otherwise, user activities are suspicious. \cite{tuorDeepLearningUnsupervised2017} proposes a stacked LSTM structure to capture the user activities in a day and adopts negative log-likelihoods of user activities as the anomalous scores to identify malicious sessions. Rather than only using the activity type, e.g., web visiting or file uploading, for insider threat detection, \cite{yuanInsiderThreatDetection2019} proposes a hierarchical neural temporal point processes model to capture both activity types and time information in a user session and then derives an anomalous score based on the differences between the predicted results and real activities in terms of types and time.

\subsubsection{Convolutional Neural Network}

Convolutional Neural Network (CNN) has achieved great success in computer vision. A typical CNN structure consists of a convolutional layer followed by a pooling layer and a fully connected layer for prediction. The convolutional and pooling layer ensures that the extracted features from inputs are rotational and positional invariant, which is a very useful property for image classification. The modern CNNs are extremely deep with tens of convolutional and pooling layers, which can achieve promising performance for image classification \cite{heDeepResidualLearning2016,krizhevskyImageNetClassificationDeep2012}. 

A recent study on insider threat detection proposes a CNN-based user authentication method by analyzing mouse bio-behavioral characteristics \cite{huInsiderThreatDetection2019}. The proposed approach represents the user mouse behaviors on a computer as an image. If an ID theft attack occurs, the user mouse behaviors will be inconsistent with the legal user. Hence, a CNN model is applied on  images generated based on the mouse behavior to identify potential insider threats.

\subsubsection{Graph Neural Network}

Graph neural network (GNN), which is able to model the relationships between nodes, has gained increasing popularity for graph analysis \cite{scarselliGraphNeuralNetwork2009,wuComprehensiveSurveyGraph2019}. A wildly-used GNN model is a graph convolutional network (GCN) that uses graph convolutional layer to extract node information. The graph convolutional layer has similar properties of a typical convolutional layer, such as local connections and shared weights, which are suitable for graph analysis. For example, the nodes in graphs are usually locally connected so that the convolutional layer is able to aggregate the feature information of a node from its neighbors. Besides GCN, graph embedding, which aims at learning low-dimensional latent representation of nodes in a network, has also attracted a lot of attention. The learned node representations can be used as features for various graph analysis tasks, such as classification, clustering, link prediction, and visualization \cite{chenTutorialNetworkEmbeddings2018}.

Recent work \cite{jiangAnomalyDetectionGraph2019} adopts a GCN model to detect insiders. Since users in an organization often make connections to each other via email or operation on the same devices, it is natural to use a graph structure to capture the inter-dependencies among users. Besides taking the adjacency matrix of structural information as input, GCN also incorporates the rich profile information about the users as the feature vectors of nodes. After applying the convolutional layers for information propagation based on the graph structure, GCN adopts the cross-entropy as the objective function to predict malicious nodes (users) in a graph. 

Inspired by graph embedding methods, research in \cite{liuLog2vecHeterogeneousGraph2019} proposes log2vec to detect malicious activities. Log2vec first constructs a heterogeneous graph by representing various activities in audit data as nodes and rich relationships among activities as edges and then train node embeddings that can encode activity relationships. Finally, by applying clustering algorithms on the node embeddings, log2vec is able to separate malicious and benign activities into different clusters and identify malicious ones.
\section{Challenges} 
\label{sec:challenge}

Although some progress has been achieved using deep learning models for insider threat detection, there are many unsolved challenges from the perspectives of characteristics of underlying data, insider threat, user expectations of detection algorithms, testbed and evaluation metrics development. In the following, we highlight the identified ten key challenges:

\begin{itemize}
    \item {\bf Extremely Unbalanced Data}. Compared with the benign activities, the malicious activities from insiders are extremely rare in real-world scenarios. Therefore, the insider threat dataset is an unbalanced dataset, which is a big challenge to train deep learning models. In general, deep learning models, which consist of tons of parameters, require large amounts of labeled data to train properly. However, it is infeasible to collect a large number of malicious insiders in reality. How to leverage the existing small samples to properly train the deep learning models is crucial to the insider threat detection task.
    
    \item {\bf Temporal Information in Attacks.} Most of the existing approaches for insider threat detection only focus on the activity type information, such as copying files to a removable disk or browsing a Web page. However, it is insufficient to detect attacks simply based on activity types conducted by users as the same activity could be either benign or malicious. A simple case is that copying files in working hours looks normal, but copying files at mid-night is suspicious. The temporal information plays an important role in analyzing user behavior to identify those malicious threats, and how to incorporate such temporal information is challenging.
    
    \item {\bf Heterogeneous Data Fusion}. Besides the temporal information, leveraging various data sources and fusing such heterogeneous data are also critical to improve the insider threat detection. For example, a user who copes files in a daily routine foresees his potential layoff and has activities of copying credential files to the removable disk in purpose. In such scenarios, considering the user profile (i.e., psychometric score) or user interaction data could help to identify potential insider threats.
    
     \item {\bf Subtle Attacks}. Currently, most of the existing work consider the insider threat detection task as the anomaly detection task, which usually models anomalous samples as out-of-distribution samples. The existing models are usually trained on samples from benign users and then applied to identify insiders that are dissimilar to observed benign samples. A threshold or anomalous score is derived to quantify the dissimilar between insiders and benign users. However, in reality, we cannot expect insiders have a significant pattern change to conduct malicious activities. In order to evade detection, insider threats are subtle and hard to notice, which means that insiders and benign users are close in the feature space. The traditional anomaly detection approaches cannot detect insiders that are close to benign users.  
    
    \item {\bf Adaptive Threats.} The insiders always improve attacking strategies to evade detection. However, the learning-based models are unable to detect new types of attacks after training. It is inefficient to train the models from scratch again when new types of attacks are observed. First, it usually needs some time to collect enough samples to train the model. More importantly, the re-training strategy cannot ensure in-time detection and prevention. Designing a model that can adaptively improve the performance of insider threat detection is an important and challenging task.
  
    \item {\bf Fine-grained Detection.} The existing deep learning based approaches usually detect malicious sessions that contain malicious activities. However, users usually conduct a large number of activities in a session. Such coarse-grained detection faces the problem that it is hard to achieve in time detection. Hence, how to identify the fine-grained malicious subsequence or the exact malicious activity is important for insider threat detection. It is also a very challenging task. This is because the information we can leverage from each activity is very limited, i.e., we only observe when and what activity is conducted by a user. Without enough information, it is hard to achieve fine-grained insider threat detection. 
    
    \item {\bf Early Detection.} Current approaches focus on insider threat detection, which means  malicious activities already occur and the significant loss is already caused to organizations. Hence, an emerging topic is how to achieve the insider threat early detection, i.e., detecting potential malicious activities ahead of they actually happen. Several approaches are proposed to defend the insider threat by using general IT security mechanisms \cite{shabtai2012survey,alneyadiSurveyDataLeakage2016}, but there is no learning-based approach to achieve early detection. Proactively identifying users who have high chances to conduct malicious activities in the near future is critical so that the organizations can conduct the intervention ahead to prevent or reduce the loss. 

    \item {\bf Explainability.} Deep learning models are usually considered as black boxes. Although deep learning can achieve promising performance in many domains, the reason why the models work is still under-exploited. When an employee is detected as an insider, it is critical to understand the reason why the model makes such predictions since employees are usually the most valuable asset in an organization. Especially, deep learning models cannot achieve 100\% of accuracy on insider threat detection. The false positive cases (misclassifying benign users as insiders) can seriously affect the loyalty of employees to the organization. Hence, the model explainability is a key to provide the insight of the model to domain-expert so that further actions can be conducted with high confidence.
    
    \item {\bf Lack of Testbed.}
    Currently, there is no real-world dataset that is publicly-available for researchers.
    Although the CERT datasets try to provide comprehensive information that are close to the high level of human realism, there is still a gap between the synthetic data and real-world scenario. 
    \begin{itemize}
        \item {\it Data Complexity.} Since the CERT dataset is a synthetic dataset, most of the activities are randomly generated with limited complexity. For example, the websites that are accessed by employees are very limited. As a result, some insider threats, such as visiting wikileak.org, can be easily identified. Meanwhile, the fine-grained user activities are randomly generated, so there is no daily routine pattern in this dataset. Furthermore, most of the activity time in the dataset are randomly generated. As a result, it is hard to leverage the temporal information to detect insider threats based on this dataset. 
        
        \item {\it Insider Threat Complexity.} The insider threat scenarios simulated in the dataset are also narrow compared with the various insider threats conducted in the real-world. The latest version of CERT dataset only consists of five scenarios. Consequently, the proposed approaches, which can achieve reasonable performance on this dataset, may not be able to achieve good performance in practice.  Meanwhile, even for the five insider threat scenarios, the difficulty of identifying these insider threats are different. 
       As shown by the ROC curves in most of the existing papers \cite{linInsiderThreatDetection2017,liuAnomalyBasedInsiderThreat2018,luInsiderThreatDetection2019,yuanInsiderThreatDetection2019}, many studies can achieve 80\% of the true positive rate with a low false negative rate. However, the false negative rate increases significantly when the true positive rate keeps increasing. It means 80\% of the insider threats in this dataset can be well detected while the rest of 20\% insider threats are hard to detect. 
    \end{itemize}

 \item {\bf Lack of Practical Evaluation Metrics.} The commonly-used classification metrics, such as  \textit{true positive rate (TPR)}, \text{false positive rate (FPR)}, \text{precision}, and \text{recall}, are adopted to evaluate performance of insider threat detection. 
Based on the TPR and FPR, an receiver operating characteristic (ROC) curve can be drawn by setting FPR and TPR as x and y axes, respectively, which indicates the trade-off between true positive and false positive. Ideally, we expect an insider threat detection algorithm can achieve TPR to be 1 and FPR to be 0. Currently, in literature, the ROC area under curve (AUC) score is widely-used to compare the performance of different detection algorithms \cite{linInsiderThreatDetection2017,liuAnomalyBasedInsiderThreat2018,luInsiderThreatDetection2019,yuanInsiderThreatDetection2019,liuLog2vecHeterogeneousGraph2019}. Another metric is the precision-recall (PR) curve, which is a plot of recall and precision as x and y axes and adopted in evaluating the unbalanced data classification. Compared to ROC-AUC, PR-AUC may be more useful to evaluate the algorithms for insider threat detection because the PR curve more focuses on the performance of classifiers on the minority class. 
However, due to the extremely small number of insiders and the corresponding malicious activities, it is unclear whether ROC-AUC or PR-AUC is practical to evaluate insider threat detection. For example, the ROC-AUC values from different detection algorithms are often close \cite{liuAnomalyBasedInsiderThreat2018,luInsiderThreatDetection2019,yuanInsiderThreatDetection2019}, which means that it is hard to identify a better model based on ROC-AUC values.


\end{itemize}
\section{Future Directions}
\label{sec:future}
The above challenges lead to several opportunities and future research directions to improve the performance of deep learning models for insider threat detection. We point out the following topics, which we believe would be promising for future research. 

\begin{itemize}
    \item {\bf Few-shot Learning based Insider Threat Detection.} 
    Few-shot learning aims at classifying samples from unknown classes given only a few labeled samples \cite{wangGeneralizingFewExamples2020}. Few-shot learning can further extend to a more rigorous setting, one-shot learning \cite{lifei-feiOneshotLearningObject2006} or zero-shot learning \cite{wangSurveyZeroShotLearning2019}, where only one or totally no labeled sample is available. Consider the extremely small number of insiders, few-shot learning is a natural fit for insider threat detection. To tackle the challenge of a few labeled samples, few-shot learning leverage the prior knowledge. Based on how to use the prior knowledge, the existing few-shot learning algorithms can be categorized into three groups, the data based approaches which augment training data by prior knowledge, the model based approaches which use the prior knowledge to constrain hypothesis space, and the algorithm based approaches which alter search strategy in hypothesis space by prior knowledge \cite{wangGeneralizingFewExamples2020}. Based on the prior knowledge we have, different few-shot learning algorithms can be developed for detecting insider threats.

    \item {\bf Self-supervised Learning based Insider Threat Detection.} Self-supervised learning aims at training a model using labels that can be easily derived from the input data rather than requiring human efforts to label the data \cite{arandjelovicLookListenLearn2017, wangLearningCorrespondenceCycleConsistency2019, asanoCriticalAnalysisSelfsupervision2019}. Self-supervised learning has achieved great success in computer vision and natural language processing. A typical self-supervised learning task in natural language processing is to pretrain a deep learning model by a language model, which is trained to predict the next word of a sentence. The task we use to pretrain the deep learning model is called ``pretext task''. After pretraining, the model can achieve great performance on the ``downstream tasks'', such as sentiment analysis, by further fine-tuning on very little data. The success of self-supervised learning is that via pretraining on the pretext tasks, the deep learning model is able to learn the salient information about the input data. In order to tackle the challenge of detecting the subtle insider threat, a potential research direction is to design the proper self-supervised tasks that can capture the difference between insiders and benign users. 
    
    \item {\bf Deep Marked Temporal Point Process based Insider Threat Detection.}
    Marked temporal point process is a powerful mathematical tool to model the observed random event patterns along time \cite{duRecurrentMarkedTemporal2016,rasmussenLectureNotesTemporal2018}. Since temporal dynamics is an important aspect of user behavior, marked temporal point process is a suitable tool to analyze the user behavior in terms of activity types and time. Recently, several deep learning based marked temporal point process models have been proposed, which usually adopt the recurrent neural network to characterize the conditional intensity function in temporal point process \cite{duRecurrentMarkedTemporal2016,liLearningTemporalPoint2018a,xiaoLearningConditionalGenerative2018}. Hence, using deep marked temporal point process models has the potential to improve the performance of insider threat detection by combining the user activity types and time information.
    
    \item{\bf Multi-model Learning based Insider Threat Detection.}  
    Because the same activity could be either benign or malicious, besides the user activity data derived from the log files, leveraging other sources is also important to improve the performance of insider threat detection. In literature, several studies investigated the performance of insider threat detection via the users' psychological data \cite{greitzerPsychosocialModelingInsider2013,almehmadiPossibilityInsiderThreat2014,hashemInsiderThreatDetection2015}, while some studies constructed user graph based on the organization hierarchy or email communication to identify the outliers \cite{jiangAnomalyDetectionGraph2019,gamachchiGraphBasedFramework2018,morianoInsiderThreatEvent2017}. However, how to combine the user activity data with the user profile data as well as the user relationship data is under-exploited and worth to explore.
    
    \item {\bf Deep Survival Analysis based Insider Threat Early Detection.}
    Survival analysis is to model the data where the outcome is the time until the occurrence of an event of interest \cite{wangMachineLearningSurvival2017}. Survival analysis is originally used in health data analysis \cite{liuEarlyPredictionDiabetes2018,luckDeepLearningPatientSpecific2017} and has been applied to many applications, such as predicting student dropout time \cite{ameriSurvivalAnalysisBased2016} or web user return time \cite{jingNeuralSurvivalRecommender2017}. 
    If we consider the time when an insider conducts a malicious activity as the event of interest, we can use the survival analysis to predict when the event (conducting a malicious activity) occurs. As a result, organizations can have early alerts about potential attacks from insiders. Recently, deep learning models are adopted to model the complex survival distributions \cite{luckDeepLearningPatientSpecific2017,chapfuwaAdversarialTimetoEventModeling2018,katzmanDeepSurvPersonalizedTreatment2018}. Hence, leveraging deep survival analysis models has a great potential to capture the user activity time information and thus achieve insider threat early detection. 

    \item {\bf Deep Bayesian Nonparametric Model for Fine-grained Insider Threat Detection.}
    In order to achieve the fine-grained insider threat detection, one potential solution is to consider activities from one user in audit data as an activity stream and apply a clustering algorithm on the stream to identify the potential malicious clusters of activities. Bayesian nonparametric models, such as Dirichlet processes, are often used for data clustering and able to generate unbounded clusters \cite{paisleyMachineLearningDirichlet2010}. The infinite nature of these models is suitable to model complicated user behavior. Recently, several Bayesian nonparametric deep generative models, which are proposed to combine the deep structure with Bayesian nonparametric \cite{zhang2018deep,nalisnickStickBreakingVariationalAutoencoders2017, goyalNonparametricVariationalAutoencoders2017}, are effective to learn rich representation based on neural networks with Bayesian methods. Leveraging the deep Bayesian nonparametric models has the potential to achieve fine-grained insider threat detection.

    \item {\bf Deep Reinforcement Learning based Insider Threat Detection.}
     Deep reinforcement learning is able to learn optimal policies for sophisticated agents in a complex environment \cite{arulkumaranDeepReinforcementLearning2017}. The advantage of deep reinforcement learning is that the policy consistently enhances its performance through reward signals. In the insider threat detection task, the insider detector can be considered as an agent in the deep reinforcement learning framework. With a properly designed reward function, the insider detector is able to keep improving the capability to identify insider attacks including the adaptive attacks. One challenge of applying the deep reinforcement learning for insider threat detection is that due to the complicated of malicious attacks, sometimes it is hard to design a good reward function. In such scenario, inverse reinforcement learning framework, whose goal is to identify a reward function automatically based on the behavior of insiders, can be further considered \cite{ohSequentialAnomalyDetection2019,ngAlgorithmsInverseReinforcement2000}. Another challenge is that deep reinforcement learning usually requires large amounts of training data that are unavailable in the insider detection task. To tackle this challenge, other machine learning paradigms, such as meta-learning or imitation learning can be further combined with the deep reinforcement learning in practice \cite{wangLearningReinforcementLearn2017,duanOneShotImitationLearning2017}. 
     Overall, although facing several challenges, deep reinforcement learning as a powerful framework has the opportunity to make a breakthrough in insider threat detection.
    
    \item {\bf Explainable Deep Learning for Insider Threat Detection.}
    Unlike some online anomaly detection tasks, such as bot detection on social media, which do not have an impact on real human beings, the insider threat detection is to identify malicious individuals, which is a high-stakes decision. Consequently, it is critical for insider threat detection models to have proper explainability even the models can achieve superior performance. Hence, how to make  prediction results understandable to human is key toward a trustworthy and reliable insider threat detection model.
    Another advantage of developing explainable deep learning models is that such models have the potential to achieve fine-grained malicious activity detection. For example, if we consider the user activity sequence in a day as a data point and each activity in the sequence as a feature, the counterfactual explanation model \cite{molnarInterpretableMachineLearning2019}, which finds a similar data point by changing some of the features for which the predicted outcome changes in a relevant way, has the potential capability to identify malicious activities from an insider activity sequence.

    \item {\bf Testbed Development.}
    To achieve insider threat detection, human actions within the monitored environment should be used as the analytical data. However, due to the privacy and confidentiality issues, the publicly available datasets in literature are very limited. Most of the recent work \cite{glasserBridgingGapPragmatic2013,linInsiderThreatDetection2017,liuAnomalyBasedInsiderThreat2018,tuorDeepLearningUnsupervised2017} adopt the CERT dataset. However, as a synthetic dataset, the user activities in CERT dataset are not complicated enough. Consequently, developing a comprehensive testbed for insider threat detection evaluation is greatly needed. 

   \item {\bf Practical Evaluation Metrics.} Due to the extremely small number of insiders and the corresponding malicious activities, the commonly-used classification metrics, such as \textit{accuracy}, \textit{F1}, \textit{ROC-AUC}, and \textit{PR-AUC} are not sufficient for evaluating the performance of insider threat detection. It is an open question which metrics are more practical and whether some new metrics need to be developed. A recent study proposes a recall based metric, called \textit{cumulative recall (CR-k)}, to evaluate the performance of algorithms on insider threat detection \cite{tuorDeepLearningUnsupervised2017}. Cumulative recall assumes that there is a daily budget $k$ to exam the top-$k$ samples with the highest malicious scores derived from the algorithms. Then, the \textit{CR-k} is defined as the sum of the recalls for all budgets up to $k$. For example, if we define $R(i)$ to be the recall with a budget of $i$, \textit{CR-k} is calculated as $R(25) + R(50) + \cdots + R(k)$. \textit{CR-k} can be considered as an approximation to an area under the recall curve. Because for insider threat detection tasks, the cost of a missed detection is substantially higher than the cost of a false positive, a recall based metric may be a suitable metric.
\end{itemize}

\section{Conclusion}
\label{sec:conclusion}
In this brief survey paper, we have reviewed various approaches in deep learning-based insider threat detection and categorized the existing approaches based on the adopted deep learning architectures. Although some progress has been achieved, the topic of using deep learning for insider threat detection is not well-exploited due to various challenges. We have discussed the challenges in this task and proposed several research directions that have the potential to advance insider threat detection based on deep learning techniques. Overall, deep learning for insider threat detection is an under-explored research topic. This survey could be extended and updated in the future as more advanced approaches are proposed.

\section*{Acknowledgments}
This work was supported in part by NSF grant 1564250.

\bibliographystyle{ACM-Reference-Format}
\bibliography{Library.bib}

\end{document}